\documentclass[lettersize,journal]{IEEEtran}
\usepackage{amsmath,amsfonts}
\usepackage{algorithmic}
\usepackage{algorithm}
\usepackage{array}
\usepackage[caption=false,font=normalsize,labelfont=sf,textfont=sf]{subfig}
\usepackage{textcomp}
\usepackage{stfloats}
\usepackage{url}
\usepackage{verbatim}
\usepackage{graphicx}
\usepackage{cite}
\begin{document}

\title{Closing the Performance and Management Gaps with Satellite Internet: Challenges,
Approaches, and Future Directions}

\author{\IEEEauthorblockN{Peng Hu\IEEEauthorrefmark{1}\IEEEauthorrefmark{2}}\\
\IEEEauthorblockA{\IEEEauthorrefmark{1}Dept. of Electrical and Computer Engineering, University of Manitoba, Canada}\\
\IEEEauthorblockA{\IEEEauthorrefmark{2}David R. Cheriton School of Computer Science, University of Waterloo, Canada}

{peng.hu@umanitoba.ca}
\thanks{}}


 \markboth{IAB Workshop on Barriers to Internet Access of Services (BIAS), January 15-17, 2024}
 {}


\maketitle



%
\section{Introduction}
Recent advancements in low-Earth orbit (LEO) satellites represented by large constellations and advanced payloads provide great promises for enabling beyond 5G and 6G telecommunications and high-quality Internet connectivity to everyone anywhere on Earth. LEO satellite networks are envisioned to bridge the urban-rural connectivity gap for the digital divide. However, the digital divide can hardly be closed by only providing connectivity to rural and remote areas without considering access equity and affordability. With these considerations, various unprecedented challenges brought by the emerging satellite Internet still need to be resolved, such as inconsistent end-to-end performance guarantees and a lack of efficient management and operations in these areas, which are referred to as ``performance gap'' and ``management gap'', respectively. This position paper will briefly discuss these gaps, approaches to addressing the gaps, and some research directions based on our recent works \cite{Hu_ComMag, Hu_MultiLayerSatNet, Hu_23cross_layer, Dhiraj_LaserISL23, Arani_HAPS_DLApproach, Atefeh_fairness_IEEEAccess21, Hu_pimrc23, Hu_SatAIOps, Sadr_ICC23, Sadr_TVT23}.

\section{Gaps Beyond Connectivity}
In this paper, the satellite Internet is considered to be enabled by advanced LEO satellite constellations, such as SpaceX's Starlink, Amazon Kuiper, Eutelsat OneWeb, and Telesat Lightspeed. If we consider the connectivity gap in the urban-rural divide, where rural and urban regions do not have access to high-speed Internet with ``a download speed of at least 50 Mbps and an upload speed of at least 10 Mbps'' \cite{StatsCan_highspeedInternet}, such a gap can be closed with the current satellited Internet. However, this does not mean we have closed the digital divide. Based on the usage gap reported in the Global Connectivity Report 2022 \cite{ ITU_report2022}, 30\% of the global population covered by a broadband network are not online ``due to lack of affordability, lack of access to a device and/or lack of awareness, skills, or purpose''. Therefore, we should aim to provide equitable, affordable, and high-quality satellite Internet access to all users, including those in rural and remote communities. To achieve this goal, we need to close the performance and management gaps as described in the following.

\subsection{Performance Gap}
The performance gap here means inconsistent end-to-end performance among satellite Internet users, including those in rural and remote areas. To examine this gap, let us focus on the fundamental throughput and latency metrics with real-world data. Let us take the latency performance in Canadian provinces, as Canada can be considered a representative country with various rural and remote communities geographically distributed across a vast landscape. Based on the available Ookla's Internet speed test results in Q2 2023 \cite{Ookla_data_url}, high-speed Internet connectivity can be achieved across the regions, but the high variance of performance metric values across regions exists, in particular for the latency performance. For example, latency results in two adjacent geographical tiles in Nunavut are 257 ms and 600 ms on fixed and mobile network connection types as defined in \cite{Ookla_methdology}, respectively. For satellite Internet, the regional performance data shown in the Starlink availability map \cite{Starlink_availability_map_url} on September 5, 2023, indicates that the variance in latency performance exists across Canada's Northern and Arctic regions. For example, the latency is 54--67 ms in Saskatchewan, 38--52 ms in British Columbia, and 44--56 ms in Quebec, while in Northwest Territories and Nunavut, it is 60--83 ms and 60--94 ms, respectively. In other regions of the world within Starlink's coverage, similar latency variances can be seen. A recent measurement of the Starlink-based satellite Internet access \cite{Pan_PIMRC23_measurement_LEOSatNet} also shows a significant variance in the round-trip time (RTT) results, fluctuating between 20 ms and 1000 ms. 

From these results, we can see that the satellite Internet can achieve low latency values but does not necessarily resolve the latency variance issue. There are many factors in the network segments of satellite Internet that may affect such latency performance. Different networking schemes used by LEO satellite network operators can create a large impact on the latency performance \cite{Zhang_Infocom22}. It is shown that end-to-end latency performance is related to the path with ground-space links and inter-satellite links (ISLs) \cite{Dhiraj_LaserISL23}, where terrestrial network (TN) facilities such as ground stations (GSs) can help reduce the ISLs used and improve consistent network performance. The current TN facilities supporting the satellite Internet in rural and remote regions are lacking, making the performance and resilience assurance provided by the satellite Internet difficult to achieve.

\subsection{Management Gap}
The lack of TN facilities in the rural and remote regions for satellite Internet exacerbates the operations and maintenance capability essential to all community and Indigenous users in rural and remote areas, where low network management costs, high responsiveness and scalability are expected. In the meanwhile, with the expansion of ground telecommunications infrastructures such as the fiber point of presence (PoP) in rural and remote communities, many existing satellite-dependent communities \cite{CRTC_SatDependent_def} are envisioned to have access to space, aerial, and ground entities, which transforms the traditional satellite-dependent community networks (SDCNs) into satellite-integrated community networks (SICNs) \cite{Hu_ComMag}, featuring an integration of heterogeneous networks and segments to provide broadband, resilient, and agile end-to-end connections. These SICNs will welcome autonomy, intelligence, and scalability in network management. However, such a transformation imposes unprecedented challenges. The complex, heterogeneous, and dynamic nature of the satellites and ground components in an integrated non-terrestrial networks (NTN) infrastructure introduces persistent operations challenges. For example, malfunctioning components, atmospheric events, and anomalous traffic on a network segment can easily degrade or disrupt network services. These challenges are collectively perceived as the management gap.




\section{Proposed Approaches \& Future Directions}

Some approaches to addressing the performance and management gaps with their research directions are discussed in the following.

\subsection{Deploying TN entities}
One way of resolving the regional performance variance on a satellite Internet is deploying additional TN entities, such as an Internet exchange point (IXP), a satellite point of presence (PoP), and an edge data center (EDC), close to a community. A satellite PoP is considered to have access to a GS. In general, IXPs can make the Internet faster and more affordable \cite{isoc2023}, and placing satellite PoPs close to communities can enhance affordability, reliability, and equity of their Internet access. An EDC can be deployed to provide consistent end-to-end latency performance for network services and applications. Similar to AWS Local Zones, which is a TN infrastructure to place compute, storage, database, and other services close to end users in large metropolitan areas, an EDC close to SDCN/SICN users for one or more communities may provide consistently low latency for the compute-or data-intensive applications for these community users. 

Designing and deploying the aforementioned TN entities for satellite Internet can lead to rich research directions. Although relevant works such as IXPs and content delivery networks (CDNs) exist in the literature, optimal deployments of TN entities for satellite Internet for different applications still demand new solutions. How to design an SICN with these TN entities for access fairness \cite{Atefeh_fairness_IEEEAccess21 }, resource orchestration, and resource allocation require future studies. In addition, how to deploy TN entities in an optimal topology to guarantee consistent performance for Internet access and digital services needs to be investigated. New measures and tools on TN entities for dynamic performance measurements are required in order to facilitate the schemes for resolving performance and management gaps.

\subsection{Multi-layer satellite networking}
LEO satellites, once deployed, are physically settled into orbital shells, which can be viewed as layers \cite{Hu_23cross_layer}.
The idea of satellite networking across different orbits can be traced in the work \cite{Nishiyama13}, where ``multi-layer satellite network'' was coined as a paradigm and a routing scheme with two-layered satellite network based on the LEO and medium-Earth orbit (MEO) was adopted to improve throughput and packet loss. Here multi-layer satellite networking (MLSN) is considered a general approach to satellite internetworking based on orbital shells and conceptual layers that can support performance-aware data transmission, intelligent space infrastructure, relay networks, efficient resource management, network operations, etc. The adoption of the layers concept can help achieve the abstraction across the dynamic inter-orbital and intra-orbital constellations of satellites operated by one or more providers. Networking may occur via multiple shells of a large constellation with LEO satellites, such as Starlink's mega-constellation \cite{Pachler21}, although the cross-shell networking is considered complex \cite{Cakaj21}. Layers of satellite networks can help formulate internetworking schemes between geostationary (GEO), MEO, and LEO satellites. Additional conceptual layers on top of these orbits can also be considered.

One research direction is devising MLSN-based schemes for performing dynamic resource management for consistent performance guarantees. These performance guarantees may be achieved across various routes consisting of multiple ISLs \cite{Dhiraj_LaserISL23} and ground-space links. Designing MLSN schemes for security countermeasures is also an important research direction due to the broaden attack surface arising from the LEO satellite constellations \cite{Giuliari2021ICARUSAL, Laursen23}. Due to the fact that the commercial LEO satellites are mostly based on proprietary technologies \cite{5GAmericas_5GNTN_2023}, the development of standardized and reliable MLSN protocols for public safety and disaster response/recovery missions calls for further research. Our recent works in \cite{Hu_MultiLayerSatNet, Hu_23cross_layer} have shown the efficiency of applying the MLSN approach to address the timing requirements and resilience assurance for message transmissions in telemetry, tracking, and control (TT\&C) missions. MLSN schemes providing consistent end-to-end performance with equitable access across areas in different scenarios need to be explored. MLSN schemes for efficient satellite network management and operations still have much room for future contributions.




\subsection{NTN-integrated networking}
NTN-integrated networking (NTN-IN) is an approach to integrating NTN entities with TN entities to support network performance guarantees, management, and services. NTN-IN is aligned with the ongoing convergence of satellite and terrestrial networks standardized by the 3rd Generation Partnership Project (3GPP). 
With the consideration of space, aerial, and ground entities, various topics need to be explored. For example, the optimal trajectory design of aerial network entities, including high-altitude platform stations (HAPS) and unmanned aerial vehicles (UAVs) \cite{Arani_HAPS_DLApproach}, joint resource allocation and entity deployment can address the common needs in various use cases. 

One research direction is the optimal placement of the NTN and TN entities for consistent end-to-end performance assurance on the satellite Internet. Another research direction is resource management in NTN-IN-based setups, such as the resource allocation schemes in a HAPS-UAV-enabled heterogeneous network \cite{Arani_HAPS_DLApproach}. In a software-defined networking (SDN) enabled LEO satellite network \cite{Papa2018}, research topics, such as joint controller and gateway placement, can be explored to maximize the network reliability. Flow setup time minimization and efficient flow table management are also active research topics due to frequent handovers and limited flow table size on small satellites. Furthermore, an NTN-IN setup based on the open radio access network (RAN), i.e., O-RAN \cite{NTN_O-RAN}, and 3GPP NTN RAN, can enable solutions to close the management gap for SICNs while the architecture design, functional split optimization, and radio resource management are open research problems.

NTN-IN can enable a broad spectrum of digital services but the design of these digital services require further studies. As NTNs and TNs provide performance-aware essential connectivity to users and devices, new digital services can be designed with the available computing and sensing capabilities on NTN entities in an NTN-IN setup. New service deployments with broad coverage to rural, remote, and hard-to-reach places can be achieved on satellite Internet. The Internet of Things (IoT) applications and services can be deployed in geographical areas that are essential to environmental monitoring, smart agriculture, and smart aquaculture. 

\subsection{Autonomous maintenance}
Equipping a satellite Internet infrastructure with autonomous capabilities such as autonomous maintenance (AM) capability \cite{Hu_ComMag} can offer ideal solutions to close the management gap. Such solutions can handle the increasing complexity of a satellite Internet consisting of various NTN/TN entities and mitigate performance degradation. In this case, AM can enhance network performance on an SDCN or SICN in supporting various applications for healthcare, education, businesses, etc. Through a data-driven architecture, generalizable anomaly identification schemes based on machine learning (ML) methods can be designed for SDCNs, SICNs and other networks using satellite Internet. As uninterrupted use of satellite Internet services is usually not guaranteed, AM can also help assure network resilience. 

AM solutions can help remove the barriers preventing users from accessing equitable and high-quality Internet connections. Bringing AM solutions to satellite Internet can help resolve the issues related to the usage gap \cite{ITU_report2022} that prevent people from using broadband networks. For example, self-diagnosing the issues and self-managing the network can reduce operating and capital expenditures, improving the access affability for users.

One research direction is designing efficient anomaly identification and mitigation methods to handle anomalous events on satellite Internet. This research includes the ML-based root cause analysis (RCA) with multivariate time-series (MTS) data from LEO satellite networks \cite{Sadr_ICC23, Sadr_TVT23}. The MTS data including onboard states, network measurements, diagnostic signals, and protocol traces can be used to devise efficient ML-based solutions. This data may also be used in threat detection caused by malicious attacks on NTN and TN segments of the satellite Internet. This research needs to consider standards-based NTN architectures to derive generalizable fault identification and localization schemes through transfer learning and federated learning with novel ML models and frameworks. The generation of high-quality open datasets on different segments and entities of the satellite Internet to facilitate the designs and benchmarks of ML-based RCA solutions is another important research topic. New measurement tools and frameworks deployable on real-world TN entities of an LEO satellite network \cite{Pan_PIMRC23_measurement_LEOSatNet} require further contributions.

From the perspective of artificial intelligence (AI) for IT operations (AIOps), developing cross-cutting AI services based on the entire life-cycle of satellites can address performance and management gaps and can enable new application domains \cite{Hu_SatAIOps}, such as space circular economy, satellite-integrated networks, green space communication, and digital twinning. 


It is worth noting that the proposed approaches may be used in combination. For example, a distributed ML model for an AM scheme may use MLSN for interaction with LEO satellite nodes. A 5G NTN network service may require an NTN-IN-based optimal joint optimization using MLSN schemes and TN entities.




\section{Conclusion}
Satellite Internet provided by advanced LEO satellite constellations is expected to be pivotal in the next-generation global telecommunications infrastructure and beyond. Beyond closing the connectivity gap in the urban-rural divide, the performance and management gaps discussed in the paper must be resolved to close the digital divide. With the proposed approaches and research directions, a future satellite Internet is expected to realize equitable, affordable and high-quality Internet access and to unlock a broad spectrum of digital services for everyone, on our way of accelerating the implementation of the Sustainable Development Goals of the United Nations.

\section*{Acknowledgments}
We acknowledge the support of the Natural Sciences and Engineering Research Council of Canada (NSERC), [funding reference number RGPIN-2022-03364].

\bibliographystyle{IEEEtran}
\bibliography{./references}


\vfill

\end{document}